\documentclass{article}
\usepackage{graphicx} % Required for inserting images
\usepackage[english]{babel}
\usepackage[utf8]{inputenc} % allow utf-8 input
\usepackage[T1]{fontenc}    % use 8-bit T1 fonts
\usepackage{hyperref}       % hyperlinks
\usepackage{url}            % simple URL typesetting
\usepackage{booktabs}       % professional-quality tables
\usepackage{amsfonts}       % blackboard math symbols
\usepackage{amsmath}
\usepackage{amssymb}
\usepackage{siunitx}
\usepackage[font=small,labelfont=bf]{caption}
\usepackage{subcaption}
\usepackage[normalem]{ulem}
\usepackage{dsfont}
\usepackage{geometry}
\usepackage{nicefrac}       % compact symbols for 1/2, etc.
\usepackage{microtype}      % microtypography
\usepackage{lipsum}		% Can be removed after putting your text content
\usepackage{xcolor}
\usepackage{graphicx}
\usepackage{algorithm}
\usepackage{tikz}
\usepackage{amsthm}

\usepackage{wrapfig}
\usepackage{csquotes}
\usepackage{braket}
\usepackage{siunitx}
\usepackage[
    backend=biber,
    style=ieee,
    sorting=none,
  ]{biblatex}
\usepackage{doi}
\usepackage{listings}
\usepackage{lettrine}
\usetikzlibrary{decorations.pathmorphing}
\AtEveryBibitem{%
  \clearfield{issn}   %Remove issn
   \clearfield{isbn}
   \clearfield{urldate}
   \clearfield{urlmonth}
   \clearfield{urlyear}
   \clearfield{urlday}
   \clearfield{month}
   \clearfield{language}
    \ifentrytype{article} {\clearfield{url}}{}
 }

\DeclareUnicodeCharacter{223C}{\ensuremath{\sim}}

\usepackage{subfiles}
\usepackage{bm}
\usepackage{authblk}
\DeclareUnicodeCharacter{223C}{\ensuremath{\sim}}

\title{Infrared markers of topological phase transitions in quantum spin Hall insulators}
\author[1]{Paolo Fachin}
\author[1]{Francesco Macheda}
\author[1,2]{Paolo Barone}
\author[1,3]{Francesco Mauri}

\affil[1]{Dipartimento di Fisica, Sapienza Università di Roma, Piazzale Aldo Moro 5, I-00185 Roma, Italy}
\affil[2]{CNR-SPIN, Area della Ricerca di Tor Vergata, Via del Fosso del Cavaliere 100, I-00133 Roma, Italy}
\affil[3]{Istituto Italiano di Tecnologia, Graphene Labs, Via Morego 30, I-16163 Genova, Italy}

\begin{document}
	
	\maketitle
	\begin{abstract}
	Using first principles techniques, we show that infrared optical response can be used to discriminate between the topological and the trivial phases of
	two-dimensional quantum spin Hall insulators (QSHI). We showcase germanene and
	jacutingaite, of recent experimental realization, as prototypical systems where
	the infrared spectrum is discontinuous across the transition, due to sudden and
	large discretized jumps of the value of Born effective charges (up to ∼2). For
	these materials, the topological transition can be induced via the application of
	an external electrostatic potential in the field-effect setup. Our results are rationalized in the framework of a low-energy Kane-Mele model
	%, and on the analysis of the vibrational modes of the systems, 
	and are robust with respect to dynamical effects which come into play when the energy gap of the material is of the same order of the infrared active phonon frequency. 
	%Thus, vibrational features in the IR absorption are markers of the topological phase transitions.  
	In the small gap QSHI germanene, due to dynamical effects, the in-plane phonon resonance in the optical conductivity shows a Fano profile with remarkable differences in the intensity and the shape between the two phases. Instead, the large gap QSHI jacutingaite presents several IR-active phonon modes whose spectral intensities drastically change between the two phases
	\end{abstract}

\section{Introduction}
Topological Quantum Spin Hall Insulators (QSHIs), firstly introduced by Kane and Mele \cite{PhysRevLett.95.226801}, present an even number of helical edge states protected by time-reversal symmetry \cite{Vanderbilt2018,HasanKanereview, Bernevig2013,doi:10.1021/acs.nanolett.9b02689,PhysRevMaterials.7.094202}. These counterpropagating modes of opposite helicity realize one-dimensional wires where the suppression of backscattering due to spin-momentum locking and time-reversal symmetry allows for dissipationless currents, with the possibility of promising technological devices such as topological field-effect transistors \cite{Gilbert2021-pc, https://doi.org/10.1002/adma.202008029,Weber2024}.
Even though the Kane-Mele model was ideated for graphene, in practice this is not a QSHI because the spin-orbit coupling inducing non trivial band inversion is too weak. Nonetheless, graphene-like monoelemental 2D honeycomb as germanene with a buckled structure \cite{bampoulisgermaneneprl2023} still realizes the Kane-Mele model, ensuring the possibility to tune topological phases thanks to the application of an orthogonal electric field \cite{Ezawa2012}.  In addition, the germanene monolayer has been shown to host a QSHI state with a sufficiently large gap such that topological properties are robust even at room temperature \cite{KlaassenJMaterChemC2024,bampoulisgermaneneprl2023,PhysRevB.109.115419}, which is essential for technological applications.  An electric-field tunable topological state, so far detected in germanene \cite{bampoulisgermaneneprl2023} and Na$_3$Bi \cite{Collins2018-cr} in a configuration where the scanning tunneling microscope (STM) allows to apply a gate field, finds promising application in the realization of the topological field-effect transistor \cite{Weber2024,doi:10.1021/acs.nanolett.1c00378}. In this device the 'ON' and 'OFF' states correspond to the presence or the absence of topologically protected ballistic transport along the edge channels of the 2D insulators. 

The quest for a large gap, stable at ambient condition, and electric-field tunable QSHI, allowing for room temperature dissipationless electronic transport, found a promising candidate in jacutingaite (Pt$_2$HgSe$_3$), a natural occurring layered mineral, first discovered in 2008 \cite{https://doi.org/10.1111/j.1365-3121.2007.00783.x} and later also artificially synthesized \cite{2012CaMin..50..431V}. In its bulk phase, jacutingaite
%The bulk system 
displays a peculiar dual topology where a topological crystalline phase is coexisting with a weak $\mathcal{Z}_2$ topological state \cite{PhysRevResearch.2.012063,PhysRevLett.124.106402,PhysRevB.102.235153}. The exfoliated monolayer is predicted to be a large-gap Kane-Mele QSHI whose topological properties are described by a buckled Kane-Mele model, analogous to the one of germanene, realized by the Hg s orbitals \cite{PhysRevLett.120.117701}. Exfoliation to few layers \cite{Kandrai2020} suggests the possibility of realizing a large-gap QSHI monolayer system, that can be even integrated in heterostructures preserving its topological properties \cite{PhysRevLett.120.117701}. 

The detection of QSHI states has been based so far on either charge transport measurements directly probing the longitudinal resistance in Hall bar geometry \cite{doi:10.1126/science.1133734, doi:10.1126/science.1148047,doi:10.1126/science.1174736,PhysRevLett.107.136603,doi:10.1126/science.aan6003,Fei2017-zh,doi:10.1021/acs.nanolett.0c01649} or discriminating conductive edge states from insulating bulk by directly imaging the local conductivity with STM measures \cite{Kim2016-bc, Tangwte2natphys2017, reisbismuthenescience2017,PhysRevB.96.041108,Deng2018-rx,Collins2018-cr, shiwte2sciadv2019, doi:10.1021/acs.nanolett.9b02444, Kandrai2020, Stuhler2020-nf,shumiyanatmat2022, Maximenko2022,Jia2022-ef,PhysRevLett.129.116802,https://doi.org/10.1002/adma.202309356, bampoulisgermaneneprl2023, PhysRevB.109.115419, KlaassenJMaterChemC2024,Xu2024-nb}. Supporting evidence for the presence of topological phases has also been proposed by comparing \emph{ab initio} band structures with single-particle spectral functions measured via
%may come also from a comparison of \emph{ab initio} predictions of the bands  with the one measure with 
angle-resolved photoemission spectroscopy (ARPES)  \cite{reisbismuthenescience2017, Tangwte2natphys2017}. %nonetheless not characterizing their topological nature directly. 
Despite not providing a direct characterization of topological properties,
%Namely, 
this technique has been used to show the topological gap reduction due to the electric field in $\textrm{Na}_3\textrm{Bi}$ \cite{Collins2018-cr}. %Nonetheless, 
Analogously, the presence of a finite local density of states (LDOS) on the edge states of a system probed by the STM measures sometimes does not provide a sufficient evidence for the existence of a topological non-trivial state as compared to the measure of ballistic edge conductivity, which in instead a robust signature of QSHI phase, but more articulate to measure. 
%As a matter of fact, 
As an illustrative example, nanoscale four-tip STM transport measurements performed on (Bi$_{0.16}$Sb$_{0.84}$)$_2$Te$_3$ showed that the observation of an edge state in the LDOS does not guarantee the presence of ballistic conductance in the edge states, the necessary condition for a QSHI state \cite{https://doi.org/10.1002/qute.202200043}. Thus, since both STM and ARPES cannot provide an indisputable proof of the presence of topological states the experimental identification of QSHI states is still a challenge motivating the seek either for improvements in the already used techniques or for alternative methods \cite{Weber2024}. Some theoretical proposals suggest plasmon based detections \cite{PhysRevLett.112.076804, PhysRevLett.119.266804} or exploiting the Ruderman-Kittel-Kasuya-Yosida interaction with magnetic impurities \cite{Duan2018} as well as on the discontinuous changes of piezoelectric response in 2D time-reversal invariant systems \cite{Yu2020,PhysRevResearch.4.L032006}
thanks to its relation with the valley Chern number or to the QSHI phase topological index $\mathcal{Z}_2$. A similar connection holds also for the Born effective charges in the prototypical Haldane and Kane-Mele models \cite{PhysRevB.110.L201405} exhibiting a discrete jump between nearly vanishing values in the topological phases and large finite ones in the trivial states. Since the Born effective charges quantify the intensity of the vibrational contribution in infrared (IR) spectra, optical spectroscopy could provide a detection method for topological phase transition. 

 In this work we  use ab initio calculations to support the feasibility of the proposal by evaluating the optical response of two real QSHI materials, germanene and jacutingaite. We here study the topological phase transition by tuning an external electric field, originating from a field-effect setup configuration. We first address germanene, a direct realization of the Kane-Mele model, and then move to the jacutingaite monolayer, where the large number of atoms enriches the picture of the Kane-Mele model, that still faithfully account for its low-energy physical properties. Finally, we test the robustness of our findings by including dynamical effects that come into play whenever phonon frequencies are in resonance with the energy gap of the material \cite{Bistoni2019,PhysRevB.103.134304}. Since the topological transition must occur across a metallic point, this resonance regime will always appear near the phase transition where the closing and reopening of the gap occurs.
% This happens, even for large band gap topological insulators near the phase transition, whenever the topological gap diminishes to the size of the phonon frequency. 
While dynamical effects are found to smoothen the discontinuous jump of the Born effective charges, vibrational contribution to the infrared spectra still presents distinct features providing a direct marker for the topological phase transition.
\section{Results}
\subsection{Theoretical framework}
\textit{Infrared spectroscopy--} The optical conductivity is used to assess the response of the system to electromagnetic fields. 
%, usually orthogonal  to the out-of-plane direction in two dimensional samples.
% For both germanene and jacutingaite the macroscopic in-plane conductivity tensor is diagonal, i.e. $\sigma_{xx}(\omega)=\sigma_{yy}(\omega),\sigma_{xy}(\omega)=\sigma_{yx}(\omega)=0$. 

%The transmittance hence simply reads
%\begin{equation}
  %  T(\omega)=\Bigg| \frac{2}{2+Z_0\sigma_{xx}(\omega)}\Bigg|^2.
    %\label{eq:transm}
%\end{equation}
In general, the conductivity is determined by two distinct contributions as
\begin{equation}
\sigma_{\alpha\beta}(\omega)=\sigma^{\textrm{el}}_{\alpha\beta}(\omega)+\sigma^{\textrm{ion}}_{\alpha\beta}(\omega),
\label{eq:optical_cond}
\end{equation}
where $\sigma^{\textrm{el}}_{\alpha\beta}(\omega)$ accounts for electronic excitations at clamped nuclei while $\sigma^{\textrm{ion}}_{\alpha\beta}(\omega)$ has resonances at the vibrational excitations of the system \cite{Bistoni2019}. We here concentrate on the expression of the ionic contribution, which reads
\begin{equation}
    {\sigma}_{\alpha\beta}^{\text{ion}}(\omega)=\frac{-i\omega}{A}\sum_{\nu} \frac{f_{\nu,\alpha}(\omega)f_{\nu,\beta}(\omega)}{\omega_\nu^2-(\omega+i\gamma_\nu/2)^2}.
    \label{eq:cond_offdiagonal}
\end{equation}
where $A$ is the area of the unit cell of the two-dimensional material, $\omega_\nu$ are the phonon frequencies with full width at half maximum $\gamma_\nu$ in the quasiparticle approximation. The oscillator strengths that determine the intensity of the infrared absorption are expressed as
\begin{equation}
    f_{\nu,\alpha}(\omega)=e\sum_{s,\delta} Z^*_{s,\alpha\delta} (\omega)\frac{e_{\nu,\delta}^s}{\sqrt{M_s}}
\end{equation}
where $M_s$ is the ionic mass, $e_{\nu,\delta}^s$ is the phonon polarization vector and $Z^*_{s,\alpha\delta}(\omega)$ are the Born effective charges, whose expression in the framework of time dependent density functional theory (TDDFT) is reported in the SI \cite{supplementary,Bistoni2019,PhysRevB.103.134304}.  The Born effective charges are the sum of a term accounting for the rigid displacement of the ionic and electronic charge ($Z^{*,\textrm{rig}}$) and an `anomalous' term describing the polarization of the electronic charge density induced by the lattice vibrations ($Z^{*,\textrm{an}}$), i.e. $Z^*=Z^{*,\textrm{rig}}+Z^{*,\textrm{an}}$. 

We now specialize to two-dimensional systems. For in-plane components, it generally holds $|Z^{*,\textrm{rig}} |\ll |Z^{*,\textrm{an}}|$ \cite{PhysRevB.110.L201405}, while generally out-of-plane components respect $|Z^{*,\textrm{an}} |\ll |Z^{*,\textrm{rig}}|$ \cite{Bistoni2019}. In the following, we will mostly concentrate our attention on the in-plane components of Born effective charges and discuss their behaviour in terms of the anomalous component. In the static limit ($\omega=0$) this is expressed in terms of the Berry curvature as \cite{PhysRevB.110.L201405}
\begin{align}
     Z^{*,\textrm{an}}_{s, \beta \alpha}=\frac{A}{(2\pi)^2} \int_{\textrm{BZ}} d^2\mathbf{k}  \Omega_{k_\alpha u_{s \beta}}(\mathbf{k}).
     \label{eq:crossder}
\end{align}

In Secs. \ref{Germanene_off_resonant} and \ref{Jacutingaite} we will study the infrared response of Germanene and Jacutingaite in the static limit through \emph{ab initio} simulation and compare to predictions performed with the Kane-Male model. Here, the approximation is that the energy of the vibrations is smaller than the typical energy gap of the material, or in other words that $\sigma^{\textrm{el}}_{\alpha\beta}(\omega_{\nu})\sim 0$  and that $Z^*_{s,\alpha\delta}(\omega_{\nu})\sim Z^*_{s,\alpha\delta}(\omega=0)$. On the contrary, in Sec. \ref{sec_dyn}  we will explicitly lift the static approximation and quantify the impact of the electronic resonances on the infrared features of the vibrational modes within the Kane-Mele model.

\textit{Kane-Mele model--} The inclusion of spin-orbit coupling in a next-nearest neighbour tight-binding model description of a 2D monoelemental honeycomb lattice naturally gives rise to topologically non trivial quantum spin hall states described by the topological index $\mathcal{Z}_2$, equal to $\pm1$ in topological phases and $0$ in the trivial ones  \cite{PhysRevLett.95.226801, Vanderbilt2018}. Originally introduced for graphene, where the weak spin-orbit coupling prevents the realization of the topological non-trivial state, the model is straightforwardly extended to a lattice with a buckled structure such as germanene and other monoelemental xenes (silicene, stanene \cite{PhysRevB.84.195430}). With the aid of Wannier functions techniques, the model has been shown to describe also the relevant low-energy physics of jacutingaite \cite{PhysRevLett.120.117701}.  

In our model Hamiltonian, we introduce an effective electric field $E^{\textrm{tot}}_{z}$ along the out-of-plane direction \cite{Ezawa2012,doi:10.7566/JPSJ.84.121003,PhysRevLett.120.117701, bampoulisgermaneneprl2023}, which breaks inversion symmetry between the sublattices. In fact, since the structure is buckled, $E^{\textrm{tot}}_{z}$ introduces an energy difference $edE^{\textrm{tot}}_{z}$ between the two sublattices, where $d$ is the half buckling height and $e$ the electronic charge, inducing an effective on-site energy that allows the system to perform transitions between the topological and the trivial phases. Our effective model Hamiltonian finally reads 
\begin{equation}
\begin{aligned}
        H_{\text{el}}=&edE^{\textrm{tot}}_z\sum_{i\rho}l_i c_{i,\rho}^\dag c_{i,\rho} -t_1 \sum_{\rho}\sum_{\langle ij \rangle}c_{i,\rho}^\dag c_{j,\rho}+i\frac{\lambda_{\textrm{SO}}}{3\sqrt{3}}\sum_{\rho\rho'}\sum_{\langle \langle ij \rangle \rangle}l_{ij} c_{i, \rho}^\dag (\sigma^\textrm{S}_z)_{\rho\rho'} c_{j, \rho'}.
\end{aligned}
\label{eq:Hamiltonian}
\end{equation}
$c_{i\rho},c_{i\rho}^\dag$ are the electronic creation and annihilation operators, where $i$ is a short-hand notation indicating both the cell and the sublattice index, $\rho=\{\uparrow,\downarrow\}$ is the spin index and $l_i=\{1,-1\}$ is the sublattice index of the site $i$. $\langle\rangle$ means sum on first nearest neighbours and $\langle\langle\rangle\rangle$ on the second ones. $d$ is the half buckling height, $t_1$ the nearest-neighbour hopping, $\lambda_{\mathrm{SO}}$ the diagonal spin orbit coupling, $\pmb{\sigma}^S$ is the vector containing Pauli matrices in the spin space, $l_{ij}=\pm1$ depending on the direction of the vector connecting the sites\cite{PhysRevLett.61.2015, Vanderbilt2018}. The system presents two direct energy gaps at $\pmb{K}$ and $\pmb{K'}$.

The trigonal symmetry of the buckled system enforced by the electric field allows for additional Rashba terms \cite{PhysRevB.91.161401} that are however negligible for the effects we will discuss. In detail, the absence of in-plane mirror symmetry allows for a further intrinsic Rashba term \cite{PhysRevB.84.195430,Ezawa2012}, that is though negligible since it vanishes at the $\pmb{K}$ and $\pmb{K'}$ points and the coupling constant is weak \cite{supplementary}. The extrinsic Rashba term usually connected with external electric fields or coupling with the substrate is also much smaller than the spin-orbit coupling leaving the unperturbed system in the topological phase and affecting the electric-field induced topological transition in a negligible way, as discussed in \cite{PhysRevB.84.195430,PhysRevLett.95.226801}.

The low energy approximation of the Kane-Mele model, obtained expanding the Hamiltonian for small quasi-momenta $\pmb{p}$ around the $\pmb{K}$ and $\pmb{K'}$ points, leads to \cite{Ezawa2012}
\begin{equation}
    H_\eta=\hbar v_\textrm{F} (p_x\sigma^\textrm{P}_x-\eta p_y\sigma^\textrm{P}_y)+(edE^{\textrm{tot}}_z-\eta\lambda_{\textrm{SO}}\sigma_z^\textrm{S})\sigma_z^\textrm{P},
\end{equation}
where $\hbar v_{\textrm{F}}=\frac{\sqrt{3}t_1 a}{2}$ is the Fermi velocity and $a$ the lattice parameter, $\sigma_z^\textrm{P}$ are the Pauli matrices in the pseudospin space describing the sublattice degree of freedom and $\eta=\pm1$ is the valley index ($\pmb{K}$ or $\pmb{K'}$ respectively).
The energy spectrum around the $\pmb{K}$ and $\pmb{K'}$ points is $E_\eta(\pmb{p})=\pm \sqrt{\hbar^2v_{\textrm{F}}^2p^2+\left(edE^{\textrm{tot}}_z-\eta\lambda_{\textrm{SO}}\right)^2}$, giving rise to a gap 
\begin{align}
\Delta_0=|2edE^{\textrm{tot}}_z-2\lambda_{\textrm{SO}}|,
\label{eq:gapDelta}
\end{align}
which closes at the critical fields $ed(E^{\textrm{tot}}_c)=\pm \lambda_{\textrm{SO}}$ marking the transition from the topological Quantum Spin Hall state ($|E^{\textrm{tot}}_z|<|E^{\textrm{tot}}_c|$) to the trivial insulating system ($|E^{\textrm{tot}}_z|>|E^{\textrm{tot}}_c|$).
The electron-phonon coupling is introduced in the tight-binding model as the linear expansion of the nearest neighbour hopping in terms of the bond length as described in Ref. \cite{PhysRevB.110.L201405}. It  enters in the low energy model as a gauge field, allowing to replace derivatives with respect to the ionic displacement with the derivatives with respect to the electronic momentum. This allows to establish a direct connection between electronic topological properties and the anomalous component of Born effective charges. Indeed, these are expressed in terms of the topological invariant $\mathcal{Z}_2$ as \cite{PhysRevB.110.L201405}
\begin{equation}
    Z^{*,\textrm{an}}_{s,xx}= (1-|\mathcal{Z}_2|)\xi \cdot l_s\textrm{sgn}(E^{\textrm{tot}}_z)\,\frac{A}{\pi},
    \label{eq:BEC_KM}
\end{equation}
where $|\mathcal{Z}_2|=1$ in the topological phases and  $|\mathcal{Z}_2|=0$ in the trivial ones,  A is the unit cell area, $\xi$ is the electron-phonon coupling parameter (see Methods [...] for numerical values).

\subsection{Germanene}\label{Germanene_off_resonant}
% The monoelemental 2D honeycomb lattice crystal composed by Germanium atoms, germanene, realizes the Kane-Mele model but with the buckled structure than enhances the spin-orbit coupling \cite{PhysRevLett.95.226801,Ezawa_2012}.
% According to the DFT simulation carried out with the Quantum Espresso suite\cite{Giannozzi_2017}, in the lowest energy configuration with lattice parameter $a=4.05\text{ \AA}$ and a buckling height of $d=0.68\text{ \AA}$, the topological gap is found be $24.3 \text{ meV}$, much more smaller than the one measured in the germanene 2D layer realized on a $GePt_2$ substrate where the topological gap is at least equal to $70 \text{ meV}$\cite{bampoulis_germanene_prl2023}, a better result for room temperature robustness of the topological properties. In addition, the experimentally measured gap lies largely above the $35 \text{ meV}$ phonon excitation energy thus justifying the off-resonant description that follows. The inclusion of the dynamical effects that becomes relevant when approaching the topological phase transition is later presented in Section \ref{sec_dyn} using the Kane-Mele model for a system compatible with the experimentally realized germanene.

\begin{figure}[h!]
    \centering
    \includegraphics[width=1\linewidth]{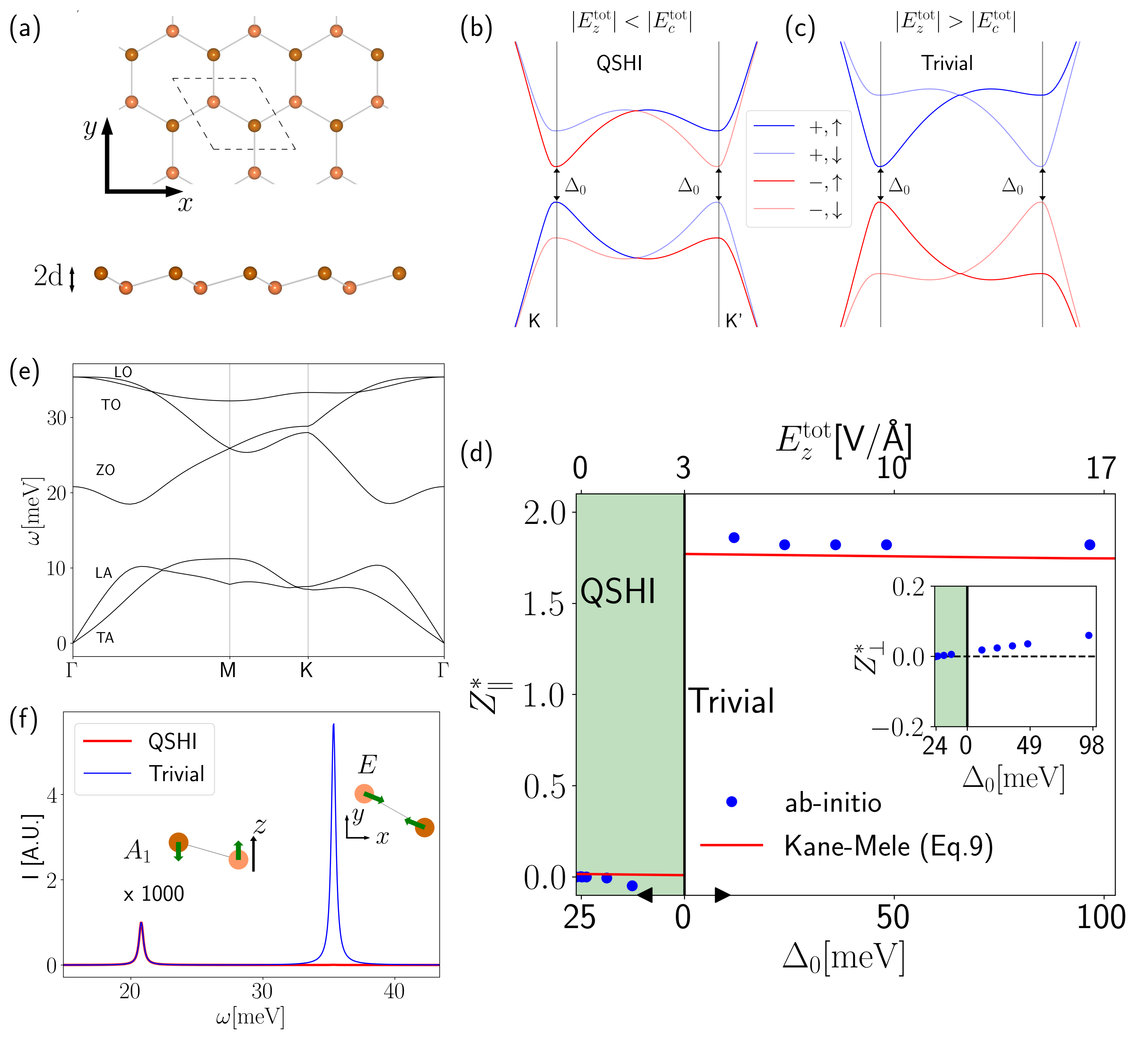}
    \caption{(a) Germanene buckled honeycomb structure, with the Cartesian reference which is chosen in this work. (b) and (c) represent the topological and trivial energetic band structure of the Kane-Mele model. The transition between the two phases is driven by the value of the external electric field $E^{\textrm{tot}}_z$, the closing of the gap happening exactly at $E^{\textrm{tot}}_c$. In (d) we show the in-plane (out-of-plane in inset) values of Born effective charges, as computed using the Kane-Mele model via Eq. \ref{eq:BEC_KM} or via \textit{ab-initio} calculations, as a function of the external electric field and the direct topological gap $\Delta_0$. Black arrows indicate increase directions for $\Delta_0$ increase. Finally, (e) is the phonon dispersion of the relevant modes at null electric field, while (f) is the simulated infrared spectrum for the $A_1$ and $E$ modes, schematically represented as insets. The intensity of the $A_1$ peak has been magnified by a factor 1000 for presentability purposes.}
    \label{fig:Ge_static}
\end{figure}

%characterization of the system, vibrations and infrared activity. Wihtout field
Freestanding germanene displays a buckled honeycomb structure with $p\bar{3}m1$ layer group (crystallographic point group $D_{3d}$), shown in Fig. \ref{fig:Ge_static}(a). 
Within our DFT calculations, the relaxed structure displays a lattice parameter of $a=4.05\text{ \AA}$, half buckling height of $d=0.34\text{ \AA}$ and a theoretical topological gap of $\Delta_0=24.3 \text{ meV}$. Despite being in agreement with previous calculations \cite{PhysRevLett.107.076802,PhysRevB.84.195430,PhysRevB.90.165431,doi:10.7566/JPSJ.84.121003}, the theoretical $\Delta_0$ is smaller than the experimental one, obtained for a sample grown on a GePt$_2$ substrate, which has a lower bound of $70 \text{ meV}$ \cite{bampoulisgermaneneprl2023}. Calculated phonon dispersion is shown in Fig. \ref{fig:Ge_static}(e). At $\Gamma$ two phonon modes are optically active: an out-of-plane $A_{1g}$ mode at $\sim 20~$meV (ZO) and an in-plane $E_g$ mode at $35~$meV (LO/TO), neither of which is IR-active. Germanium atoms form two inversion-partner sublattices arising from Wyckoff position (WP) $2d$ in the unit cell. The local $3m$ site-symmetries constraint the Born effective charge tensor to be diagonal, with two independent components $Z^*_{s,xx}=Z^*_{s,yy}$ and $Z^*_{s,zz}$, while inversion symmetry implies that the effective charges of the two Ge atoms in the cell are the same, i.e., $Z^*_{s,xx} =Z^*_{\parallel}$ and $Z^*_{s,zz}= Z^*_{\perp}$. Since the charge-neutrality condition imposes $\sum_s Z^*_{s,\alpha\beta}=0$, both components $Z^*_{\parallel}$ and $Z^*_{\perp}$ are zero, consistently with the absence of IR-active modes.

% With field
An external electric field applied perpendicularly to the monolayer plane tunes a topological transition marked by the closure of the inverted gap at a critical field $E^{\textrm{tot}}_c$, above which the trivial gap opens again, as depicted in Fig. \ref{fig:Ge_static}(b),(c). Phonon frequencies are insensitive to the topological transition, and they are weakly modified by the applied field (see section 5 SI\cite{supplementary}). On the other hand, the orthogonal electric field breaks spatial inversion, reducing the symmetry to the $p3m1$ layer group (point group $C_{3v}$). The Wyckoff position $2d$ is accordingly split to $1b+1c$, making the two Ge atoms inequivalent but preserving the local $3m$ site symmetries. It follows that Born effective charge components  are allowed to acquire a non-zero value at finite applied fields, being opposite on inequivalent Ge atoms because of the charge-neutrality condition, $Z^*_{s,xx}=l_s Z^*_\parallel$ and  $Z^*_{s,zz}=l_s Z^*_\perp$. Consistently, both optical modes become IR-active, and they are labeled as $A_1$ and $E$ irreducible representations of the $C_{3v}$ point group. The evolution of effective-charge components calculated from first principles as a function of the applied field and across the topological transition is shown in Fig. \ref{fig:Ge_static}(d). The out-of-plane component $Z^*_\perp$ displays a continuous linear increase with the applied field and it is insensitive to the topological transition, as shown in the inset of Fig. \ref{fig:Ge_static}(d). On the other hand, the in-plane $Z^*_\parallel$ component remains substantially zero in the QSHI phase, displaying only a slight deviation before undergoing a large enhancement at the critical field that induces the topological transition to the trivial phase. The topological origin of this jump can be appreciated by comparing the first-principles result with the prediction, Eq. (\ref{eq:BEC_KM}), of the Kane-Mele model for germanene, shown as a red line in Fig. \ref{fig:Ge_static}(d). The excellent agreement between first-principles and model results can be attributed to the reliability of the low-energy Kane-Mele model in capturing the (dominant) topological contribution to the Born effective charges, that arises from the neighbourhoods of the Dirac points. Even though analogous results are obtained for two isostructural and isoelectronic Silicene and Stanene (section 2 of SI \cite{supplementary}), in general further trivial contributions from other regions of the Brillouin zone can be expected to affect quantitatively the total Born effective charges, as exemplified by Jacutingaite discussed in the next section.

We are now in a position to assess the IR vibrational contribution to the optical conductivity of Germanene. In Fig. \ref{fig:Ge_static}(f) we show the IR spectra computed for two values of the electric field $E^{1}_z=0.6 \textrm{ V/nm}$ and $E^{2}_z=5.1 \textrm{ V/nm}$ corresponding to a gap $19 \mathrm{ meV}$ and $24 \mathrm{ meV}$ respectively, placing the system in the QSHI and trivial phase, respectively. In both cases, the vibrational IR spectrum is expected to display two peaks corresponding to the IR-active out-of-plane $A_1$ and in-plane $E$ modes. Since $Z^*_{\perp}<Z^*_{\parallel}$, we enhanced the $A_1$ peak by a factor 1000 for illustration purposes, and we normalized both spectra to such intensity. The (rescaled) two-peak structure is clearly visible in the trivial phase, where the $E$ peak is the most prominent feature of the IR spectrum. Instead, the spectrum in the QSHI phase only displays the $A_1$ peak, while the $E$ peak practically disappears.

%, providing a direct measurable marker of the topological transition.
%

\subsection{Jacutingaite}\label{Jacutingaite}
\begin{figure}[h!]
    \centering
    
        \includegraphics[width=1\linewidth]{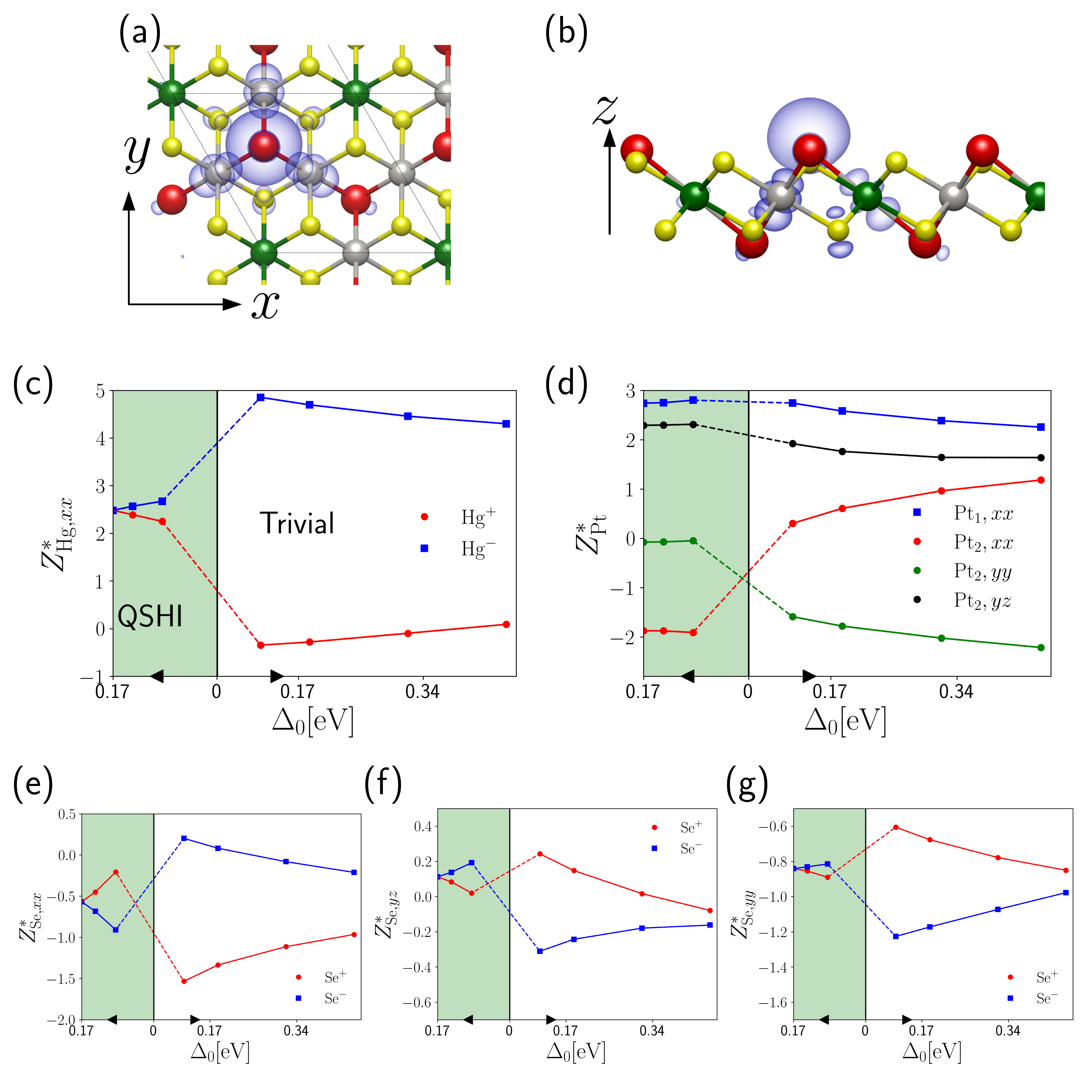}
    \caption{(a) Top and (b) side view of Jacutingaite. The colour code for the atoms is: Hg red, Pt$_1$ green, Pt$_2$ gray, Se yellow. We label the Hg atoms that are towards more positive(negative) $z$ as Hg$^{+}$(Hg$^{-}$), and same for Se$^{+}$(Se$^{-}$). Maximally localized Wannier functions built from the low-energy  spectrum are represented as isocontours. In the panels below we plot relevant components of Born effective charges as a function of the direct topological gap at $K$ point $\Delta_0$, varied applying an external orthogonal electric field, for (c) Hg$^{\pm}$, (d) Pt$_{2}$, (e,f,g) Se$^{\pm}$. Black arrows indicate increase directions for $\Delta_0$. As described in the text, Hg$^{\pm}$ strongly resemble the results of the Kane-Mele model, as per Eq. \ref{eq:BEC_rat}. On the contrary, the behaviour of Se$^{\pm}$ and Pt$_2$ atoms is more complicated, even if jumps (or, rather, components exchanges) are still rationalized in terms of the topological transition, as described in the text. Dashed lines across the transition are intended as a guide to the eye whereas the continuos ones describe the trends in each phase.}
    \label{fig:ZHg}
\end{figure}

The monolayer of jacutingaite (Pt$_2$HgSe$_3$) is a QSHI with a large indirect band gap between the K and M points of $\sim 0.15 \text{ eV}$, and a comparable topological direct gap of $\sim 0.17 \text{ eV}$ at the K point\cite{PhysRevLett.120.117701}. Its structure belongs to the same layer group $p\bar{3}m1$ of germanene or buckled graphene, the Hg atoms occupying the same $2d$ Wyckoff positions of Ge in the buckled honeycomb lattice [Fig. \ref{fig:ZHg}(a),(b)]. The low-energy physics of the topological gap is well described by an effective two-band model analogous to a Kane-Mele model \cite{PhysRevLett.120.117701}, that can be constructed in a basis of maximally localized Wannier functions (MLWFs) \cite{RevModPhys.84.1419}. The MLWFs contributing to the low-energy effective model are centered on Hg atoms, i.e. on the sites of the buckled honeycomb lattice, and they display a dominant Hg $s$ character, with weaker but non-negligible hybridization with nearest-neighbor Pt atoms bridging Hg [gray-coloured in Fig. \ref{fig:ZHg}(a),(b)] and with surrounding Se atoms. Given the stringent analogy with germanene, the topological phases of jacutingaite can be tuned by means of an external orthogonal electric field, that also in this case lowers the symmetry of the system to $p3m1$. Differently from germanene, the electric field induces a strong rearrangement of the ionic positions inside the unit cell. The ionic distorsion is responsible of a relevant reduction of the critical field for the topological transition and, as discussed in \cite{PhysRevLett.120.117701}, it represent by itself a mechanism to drive the system to the trivial state. 

Being described by an effective Kane-Mele model, the contributions to the in-plane Born effective charges coming from the low energy sector of the Hamiltonian, and therefore due to the orbitals entering in the topological band inversion mechanism, are expected to present a discontinuous jump across the topological phase transition. The number of independent components of the Born effective charge tensors is determined by the site-symmetry of the representative of each inequivalent atomic species,as discussed in the Methods section. In Figure \ref{fig:ZHg} we consider the only independent in-plane component of the diagonal Born effective charges tensor of Hg and Pt$_1$ and the non zero in-plane components of the representative Pt$_2$ and Se atoms for which the twofold rotation axis is parallel to $x$.

When the symmetry is lowered to $p3m1$, previously inversion-related Hg and Se atoms become inequivalent. We will distinguish such atoms with the labels Hg$^{\pm}$ and Se$^{\pm}$, where the $+$ means that the atom is above the plane of system and accordingly for the $-$, as shown in Fig. \ref{fig:ZHg}(b). On the other hand, inversion-symmetry breaking does not modify the local site-symmetries and the form of the Born effective charges tensors on all atoms, though the values of the components for Hg$^{+}$ (Se$^{+}$) and Hg$^{-}$ (Se$^{-}$) will now be different.

The Born effective charges computed \textit{ab initio} and the evolution through the topological transition of selected components are displayed in Fig. \ref{fig:ZHg}. Coherently with the above analysis, the in-plane component of the Born effective charge for Hg experiences a discontinuous jump across the topological transition, as shown in Fig. \ref{fig:ZHg}(c). For  Pt$_1$, not taking part to the Kane-Mele effective model, no such discontinuity appears [Fig. \ref{fig:ZHg}(d)]. Interestingly, some components of the effective charges of Pt$_2$ and Se also display discontinuities, shown in Fig. \ref{fig:ZHg}(e)-(g).

Notice that, differently from germanene, Born effective charges are not zero at null external field. This is because the low-energy sector of the electronic spectrum only contributes to a portion of the Born effective charges value, and therefore only such contribution displays a relevant jump associated with the topological phase transition, while the reminder has a trivial (and negligible in first instance) behaviour with the applied field. In other words, the \textit{ab-initio} behaviour of Born effective charges of jacutingaite as a function of $E^{\textrm{tot}}_z$ in the studied range can be rationalized via the following expression:
\begin{align}
Z^*_{s, \alpha\beta}(E^{\textrm{tot}}_z)\sim Z^*_{s, \alpha\beta}(0)+ \delta Z^{*}_{s,\alpha\beta} \theta(E_z^{\textrm{tot}}-E_c),
\label{eq:BEC_rat}
\end{align}
where $\delta Z^{*}_{s,\alpha\beta}$ is the topological jump, different for every atom and component, with the only general rule that $\delta Z^*_{s,z\alpha}=0$ where $\alpha=x,y,z$ since the out of plane induced polarization is not related to electronic topological effects. For example, $\delta Z^{*}_{\textrm{Hg}^{\pm},xx}$ clearly follows the behaviour of Eq. \ref{eq:BEC_KM}, with $l_s=\pm 1$ referring to the two inequivalent Hg$^\pm$ atoms, as expected from the dominant mercury character of the MLWFs realizing the effective Kane-Mele model. In particular, the magnitude of the discontinuous jump for Hg$^\pm$ is $\vert\delta Z^{*}_{\textrm{Hg}^{\pm},xx}\vert\sim 2.5$.

For Se$^{\pm}$ and Pt$_2$ the situation is different. In the first case, across the topological transition there is an apparent exchange of components between the partners Se$^{\pm}$. Concerning Pt$_2$, where some components display an almost trivial behaviour while others display large jumps, an analogous exchange of contributions seems to happen, but for different Cartesian components on the same atom. 
Both behaviours are reasonably traceable to the exchange of conduction and valence wavefunction occurring at the topological phase transition rather than a jump in the form of Eq. \ref{eq:BEC_KM}. In fact, Se$^{\pm}$ and Pt$_2$ orbitals enter in the determination of MLWFs, but in a weaker fashion with respect to Hg. Therefore, the proper topological jump of Eq. \ref{eq:BEC_KM} is expected to be much weaker. Nonetheless, across the topological transition some components of Se$^{\pm}$ and Pt$_2$ orbitals are energetically swapped between valence and conduction bands, as well as their contribution to effective charges via the exchange of electronic wavefunctions, ending up in exchanges of components. Finally, we notice that the trivial behaviour of Se effective charges under electric field is not fully negligible as for Eq. \ref{eq:BEC_rat}, and discussed in the section 3 of the SI \cite{supplementary}.

We turn now to analyse the vibrational IR spectrum of jacutingaite, focusing on in-plane phonon modes that are expected to display signatures of the topological transition. Even in the absence of applied field, among the 33 optical modes at $\Gamma$ listed in Supplementary Table 5 \cite{supplementary}, seven in-plane phonons belong to the IR-active $E_u$ representation, consistently with the non-zero Born effective charges in the $p\bar{3}m1$ phase of jacutingaite. Four other optically active in-plane $E_g$ modes are instead Raman-active, but they can show IR activity when the inversion symmetry is broken by the external electric field. The frequencies are quite close to the one of the bulk sample, where the \emph{ab initio} computation is in good agreement with Raman experiments \cite{10.1063/5.0053171, Longuinhos2021, https://doi.org/10.1002/jrs.5764}. Although phonon frequencies show a weak dependence on the applied field (see Section 4 of SI \cite{supplementary}), the strength of their coupling with IR radiation is significantly affected by the topology of the underlying electronic structure.

\begin{figure}[h!]
    \centering
    \includegraphics[width=0.9\linewidth]{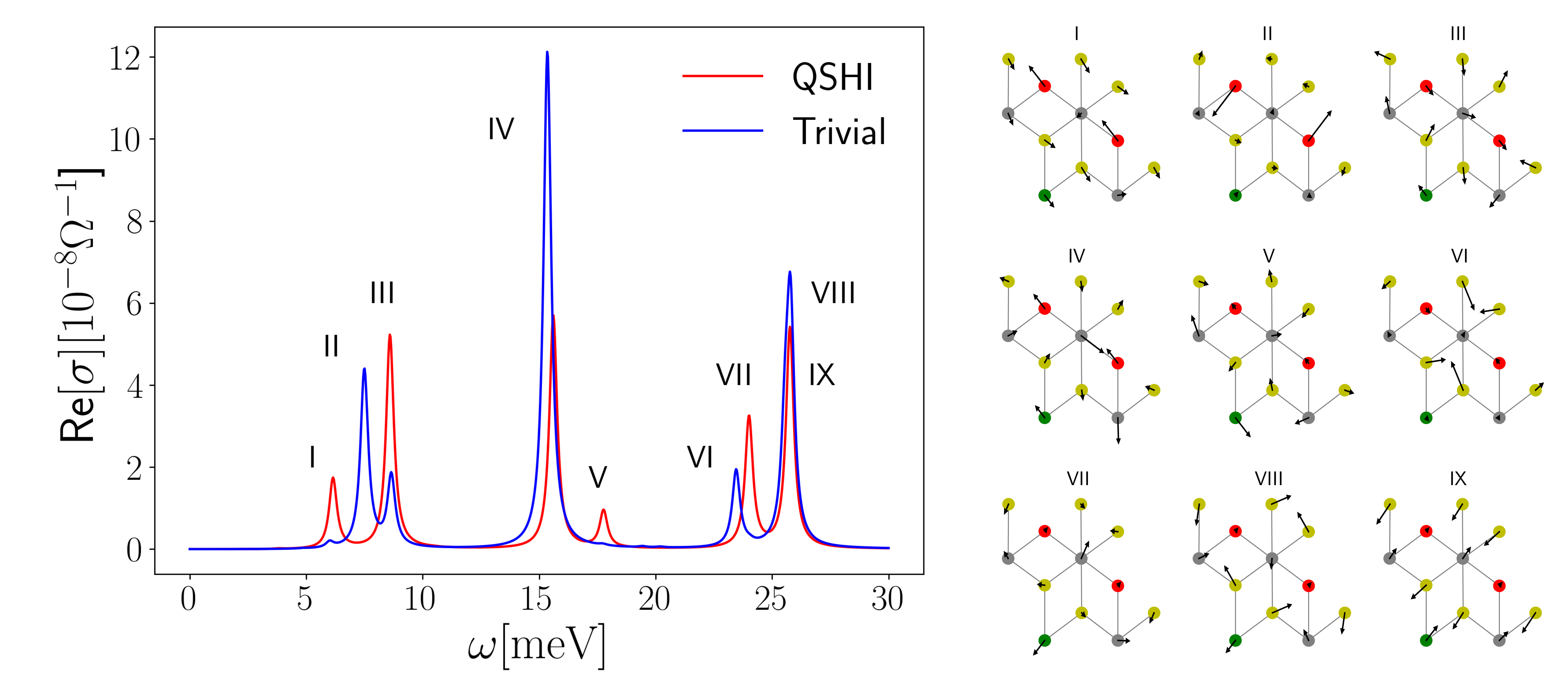}
    \caption{Left panel: infrared optical spectrum of jacutingaite, for both the QSHI and the trivial phases, for gaps at the $\pmb{K}$ point of $\Delta_1=0.15 \textrm{ eV}$ and $\Delta_2=0.18 \textrm{ eV}$. Linewidth have been multiplied by an arbitrary factor for representation purposes. 9 distinct phonons are identified and labelled with Roman numerals; the corresponding in-plane projections of the atomic displacements are represented in the right panel (the colour code for atomic species is the same as for Fig. \ref{fig:ZHg}). As discussed in the text, the mode II is the clear analogue of the $E$ mode of germanene, with the Hg$^{\pm}$ atoms playing the same role of the Ge one. The behaviour of the other modes is explained in the text.}
    \label{fig:spectra_cfr}
\end{figure}

The computed spectrum at finite fields for the QSHI phase with a topological gap of $\Delta_1=0.15 \textrm{ eV}$ is compared in Fig. \ref{fig:spectra_cfr} with that of the trivial phase with a comparable gap $\Delta_2=0.18 \textrm{ eV}$. Out of the 11 IR-active modes, only nine are found to significantly contribute to the optical response, which we label by Roman numerals. The most interesting ones are the vibrational modes II and VI, both deriving from the $E_g$ Raman-active modes of inversion-symmetric jacutingaite at zero field. Mode II is due to Hg atoms moving in opposite phase, and it is the direct analogue of the $E_g\rightarrow E$ mode of germanene, coherently being infrared inactive in absence of external field and displaying a huge intensity jump across the topological phase transition. The behaviour of mode VI can be similarly understood as due to the out-of-phase in-plane motion of Se atoms, giving rise to a negligible though finite IR intensity that undergoes a sizeable enhancement through the topological transition.
% at odds with the other two $E_g$-derived modes whose contribution to the conductivity remains negligible for all considered values of the field (thus not discussed here, for details see SI Tables ...). 
Finally, also the intensity of the seven modes that are IR-active at zero field changes relevantly as the system evolves from the QSHI to the trivial phase. Among these, only mode I is significantly contributed by Hg atoms, that move in phase and with the same amplitude in absence of the field (see Fig. \ref{fig:spectra_cfr}); with increasing field, a growing unbalance between the Hg$^{\pm}$ atomic displacements enhances the dependence on the topological contribution $\delta Z_{\textrm{Hg}^{\pm},xx}$ of the related oscillator strengths, ultimately leading to a discontinuous drop of its IR intensity at the topological transition. Similar drops are observed for modes III, V and VII, while mode IV and quasi-degenerate modes VIII, IX display increasing intensity in the trivial phase. Despite displaying distinct and detectable IR spectral features related to the topological phase transition, all such changes are contributed mostly by Pt atoms, and cannot thus be rationalized within the results of the effective Kane-Mele model.

\subsection{Dynamical effects}\label{sec_dyn}
If the topological gap $\Delta_0$ is of the order of magnitude of a relevant infrared active phonon frequency $\omega_{\textrm{ph}}$, the phonon frequency become resonant with electronic excitations and therefore the Born effective charges cannot any more be approximated by their static value. This is for example the case for germanene, if the gap is taken as its theoretical DFT value. Nonetheless, 
also for large topological gap insulators such as jacutingaite, there will always be a regime where $\Delta_0 \sim \omega_{\textrm{ph}}$ near the topological transition. In this case, the vibrational IR response of the system is quantified by the complex valued frequency dependent dynamical effective charges \cite{Bistoni2019,PhysRevB.103.134304}. The interplay between the electronic and phonon excitation produces frequency dependent Fano lineshapes on top of the electronic excitation spectrum in the optical spectra, rather than off resonant Lorentzian lineshapes, which qualitatively and quantitatively change the shape of the infrared spectrum, as predicted and seen experimentally in graphene systems \cite{PhysRevB.86.115439}.

Since dynamical effects are relevant when the electronic gap is small, they are well described by the low-energy Kane-Mele model. The reliability of the model is further verified by comparing the Kane-Mele and the ab initio static electric susceptibility reflecting the effect of electronic excitations (shown in Supplementary Fig. 12). It is then easy to obtain analytical expressions for $\sigma^{\textrm{el}}(\omega)$, as well as for the diagonal in-plane-components $Z^*_{s,xx}(\omega_{\textrm{ph}})$, as shown in the section 5 of the SI \cite{supplementary}. Notice that the relation $Z^*_{s,xx}(\omega_{\textrm{ph}})=l_sZ^*(\omega_{\textrm{ph}})$ still holds even in the presence of dynamical effects if, as assumed in this work following Ref. \cite{PhysRevB.110.L201405}, the electron-phonon coupling depends only on the relative distance between two atoms. 
The conductivity tensor is also diagonal for $p\bar{3}m1$ and $p{3}m1$ symmetries, with $\sigma_{xx}=\sigma_{yy}$; in the following we refer to the in-plane component as $\sigma$. 

We now sum up the most important results. Defining $z=(\omega-\omega_{\textrm{ph}})/\gamma_{\textrm{ph}}$, the ionic contribution to the optical conductivity is described by a Fano function \cite{PhysRev.124.1866}
\begin{equation}
    {\sigma}^{\text{ion}}(\omega)=P\frac{q^2-1+2qz}{(1+q^2)(1+z^2)}+iP \frac{(q^2-1)z-2q}{(z^2+1)(1+q^2)},
\end{equation}
where
\begin{equation}
    P=\frac{\sigma_0 \bar{\omega}}{\gamma_{\textrm{ph}}} |Z(\omega_{\textrm{ph}})|^2 , \quad
    q=-\frac{\text{Re}[Z^*(\omega_{\textrm{ph}})]}{\text{Im}[Z^*(\omega_{\textrm{ph}})]},
\end{equation}
are the phonon strength and the Fano asymmetry parameters, respectively. Depending on the asymmetry parameter, the Fano function can describe a variety of different profiles ranging from a positive Lorentzian peak ($q \to -\infty$, when $\textrm{Im}Z^*\rightarrow 0$ ) to highly antisymmetric dispersive profiles ($q \to -1$) and even negative peaks ($q=0$)  \cite{PhysRevB.86.115439}. For germanene, $\hbar \bar{\omega}=\frac{8 \hbar^2}{M_{\textrm{Ge}} A}=0.032 \text{ meV}$, while $\sigma_0=\frac{\pi e^2}{2h}$ is the universal conductivity, corresponding to the one of clean graphene. The effects of the electric field on $\omega_\textrm{ph}$ are taken into account by computing the variation of the static frequency from \emph{ab-initio} calculations, while dynamical effects on the phonon frequency are computed analytically within the low energy model (see section 5 SI\cite{supplementary}). The latter are small compared with the first ones, causing only a slightly enhancement at the resonance between the phonon frequency and the gap. $\gamma_{\textrm{ph}}=\gamma_{\textrm{ph-ph}}+\gamma_{\textrm{e-ph}}$ is the full width at half maximum determined by the phonon-phonon scattering, taken from experimental measures on bulk Germanium and assumed independent from electronic resonances $\gamma_{\textrm{ph-ph}}=0.07 \text{ meV} $\cite{PhysRevB.57.1348}, and the electron-phonon coupling contribution $\gamma_{\textrm{e-ph}}$, computed including dynamical effects in the Kane-Mele model as discussed in section 5 of the SI \cite{supplementary}. The phonon linewidth is significantly enhanced by the electronic resonance, almost doubling its value. 

\begin{figure*}
    \centering
    \includegraphics[width=1\linewidth]{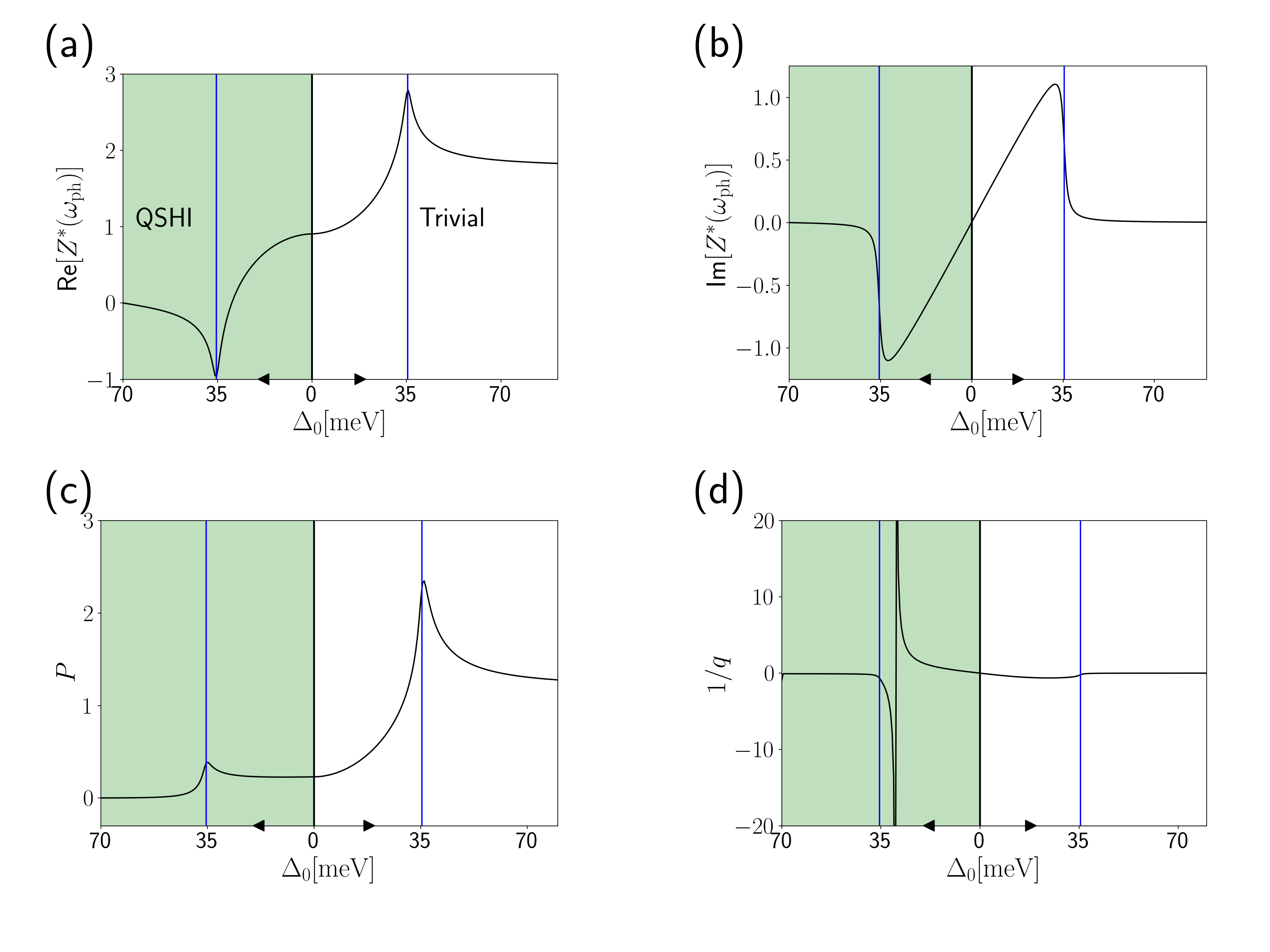}
    \caption{(a) Real and (b) imaginary parts of the dynamical Born effective charges for the Kane-Mele model across the QSHI and trivial phases as a function of the direct gap at $K$ point $\Delta_0$ varied applying an external orthogonal electric field. The topological unperturbed gap is taken from the one experimental Germanene $70\textrm{ meV}$, whereas the static phonon frequency of $\omega_{\textrm{ph}}=35.5\text{ meV}$ as well as the electron-phonon coupling parameters are predicted by \textit{ab-initio} calculations. The related phonon strength and Fano asymmetry parameters for the ionic conductivity are represented in the panels (c) and (d). Black arrows indicate increase directions for $\Delta_0$ while the vertical blue lines correspond to the static phonon frequency energy marking the condition $\Delta_0=\omega_{\textrm{ph}}$ }
    \label{fig:parameters}
\end{figure*}

The complex valued dynamical effective charges, the phonon strength $P$ and the Fano asymmetry $q$ parameters are plotted in the Figure \ref{fig:parameters} for a Kane-Mele model with $\Delta_0=70 \text{meV}$ at null external field, as compatible with the experimentally measured one in germanene \cite{bampoulisgermaneneprl2023}, and a static phonon frequency of $\omega_{\textrm{ph}}=35.5 \text{ meV}$ at null external field as predicted by \textit{ab-initio} calculations. 
The topological transition affects the Born effective charges and the parameters $P$ and $q$ in a different way with respect to the static case. Most notably, the discontinuous jumps of Eq. \ref{eq:BEC_KM} are smoothed by dynamical effects, and Born effective charges acquire a relevant imaginary part in the resonant condition with an opposite sign in the two topological phases %QSHI phase.

\begin{figure}
\centering
    \includegraphics[width=0.99\linewidth]{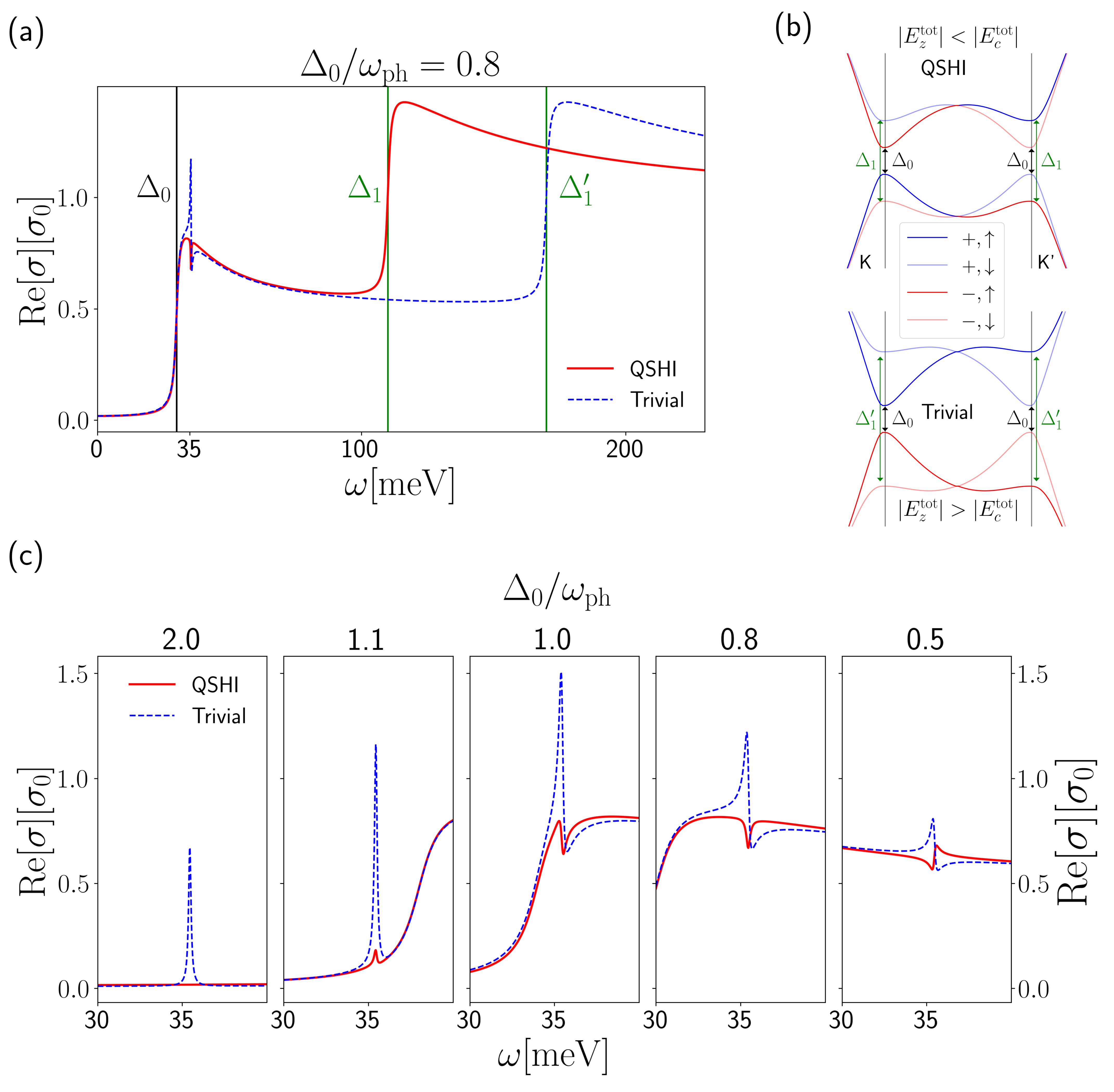}
    \caption{(a) Real part of the optical conductivity of the Kane-Mele model, taking into account dynamical effects on both the electronic and ionic contributions, in the QSHI and the trivial phases for the same direct gap $\Delta_0=0.8 \omega_{\textrm{ph}}$ , where the static phonon frequency of $\omega_{\textrm{ph}}=35.5\text{ meV}$ is obtained from \emph{ab initio} calculation.  The relevant electronic transitions marked by the vertical black (direct gap transition between bands undergoing topological band inversion) and green lines (transition between the higher and lower energy bands, not affected by topological effects) are described in the nergy bands in the panel (b).  The conductivity is expressed in units of the universal conductivity $\sigma_0= e^2/4\hbar$. (c) Ionic contribution to the real part of the optical conductivity  as a function of the ratio $\Delta_{0}/\omega_{\mathrm{ph}}$, where the direct gap at the $K$ point is varied by mean of an external orthogonal electric field. Where the topological gap is much larger than the phonon frequencies, we recover the same results as for the static treatment (i.e. disappearence of the phonon feature in the QSHI phase). When the topological gap is comparable or smaller than the phonon frequencies, dynamical effects kick in, and the phonon is visible even in the QSHI phase. Nonetheless, the two phases are still very clearly distinguishable due to the different shape of the Fano profile.}
    \label{fig:conductivity}    
\end{figure}

We report in Fig. \ref{fig:conductivity} (a) the real part of the optical conductivity including both electronic and phonon features (see Eq. \eqref{eq:optical_cond}), where the relevant electronic transitions are depicted in Fig. \ref{fig:conductivity} (b). We also report the ionic contribution to the real part of the optical conductivity, in Fig. \ref{fig:conductivity} (c), for several values of $\Delta_0/\omega_{\textrm{ph}}$. In the off-resonance condition $\Delta_0/\omega_{\textrm{ph}}\sim 2.0$ the phonon spectral feature is visible only in the trivial phase while vanishing in the topological one, as already discussed in Section \ref{Germanene_off_resonant}. Approaching the resonance $\Delta_0/\omega_{\textrm{ph}}\sim 1$ the phonon contribution becomes non vanishing also in the QSHI phase but with markedly different spectral features from the trivial phase. For phonon energies larger than the gap, the Fano lineshapes describing phonon resonances between the two phases differ qualitatively, going from positive to negative peaks or having opposite dispersive behaviour. Even tough the intensity difference becomes not so large as in the off-resonant condition, 
the lineshape corresponding to different topological phases with the same gap are clearly discernible.

\clearpage
\section{Discussion}
In this work, we showed that infrared response can be used as a marker to characterize topological transitions on promising and already experimentally studied materials, such as germanene and jacutingaite. We confirmed showed via first principles calculations that the Born effective charges display sudden finite jumps across the topological phase transitions, in agreement with predictions based on a low-energy Kane-Mele model. Importantly, such jumps are very large, of order 2 for germanene and up to 2.5 for the Hg atoms of jacutingaite, causing a clear cut change in the infrared vibrational spectrum across the transition. While in Germanene the vibrational resonances almost disappear, in Jacutingaite large intensities are preserved in the markedly different spectral profiles due to the different topological states, a change that can be detected by experimental measures with the same sensitivity. We also showed that our conclusions are robust with respect to dynamical mechanisms, that become relevant when the energy gap is of the same order of the frequency of the infrared active phonon. Our results pave the way for the determination of the topological properties of materials via the use of optical spectroscopy techniques.
\subsection*{Acknowledgement }
We acknowledge the MORE-TEM ERC-SYN project, Grant Agreement No. 951215. We acknowledge the EuroHPC Joint Undertaking for awarding this project access to the EuroHPC supercomputer LUMI, hosted by CSC (Finland) and the LUMI consortium through a EuroHPC Regular Access call. 
\section{Materials and Methods}
\textit{Ab initio parameters---} First principles calculations were performed using Quantum ESPRESSO\cite{Giannozzi2017} with Perdew-Burke-Ernzerhof (GGA)  exchange functional \cite{PhysRevLett.77.3865} using Pseudo Dojo full relativistic Optimized norm-conserving Vanderbilt pseudo-potentials \cite{VANSETTEN201839,PhysRevB.88.085117} with a plane wave cutoff on the wavefunctions of 80 Ry for Germanene, Stanene and Silicene and 60 Ry for Jacutingaite. The convergence threshold for the phonon calculation is set at $10^{-18}$ (corresponding to the input word tr2\_ph in Quantum Espresso). Since Germanene, Stanene and Silicene have a small gap,  Brillouin zone sampling is carried out using a telescopic grid with an elevated density around the $\textrm{K}$ and $\textrm{K'}$ points to achieve convergence, according to the procedure described in the Appendix B of \cite{PhysRevB.109.075420} with the following parameters: for Germanene and Stanene $\mathcal{N}=50,\, l=5, \,L=9,\, p=8$ corresponding to 481 irreducible points, for Silicene $\mathcal{N}=50,\, l=11, \,L=6,\, p=8$ having 3386 irreducible points. For Jacutingaite, by virtue of the larger gap, a 18x18x1 Monkhorst–Pack mesh \cite{PhysRevB.13.5188} is sufficient to achieve convergence. An interlayer distance larger than $19 \text{\AA} $ is used to separate the monolayers of Germanene, Stanene and Silicene and of $30 \text{\AA} $ for the Jacutingaite monolayer. The $2\times 2$ maximally localized wannier function Hamiltonian describing the valence and conduction bands of jacutingaite is performed using WANNIER90 \cite{Pizzi2020}. In all our calculation we implement the 2D cutoff technique with the electric field introduced in field-effect transistor setup, in the double gate configuration \cite{PhysRevB.96.075448}. The charged plates are placed symmetric with respect to the plane of the 2D system at $z=\pm 0.221$ in crystalline units in all the studied systems. The generated capacitor-like potential increases (or decreases) linearly until the potential barriers of height $E=34 \text{ eV}$ placed at $z=\pm 0.22$ in crystalline units. 
All the  structures are relaxed with a threshold of at least $7\cdot 10^{-7} \text{eV}$ for the energy and $7\cdot 10^{-6} \text{eV/\AA}$ for the forces. 

\textit{Low energy model parameters---}The model parameters of Germanene, as well as for Stanene and Silicene, are obtained from the \emph{ab initio} simulations. 

We use Eq. \ref{eq:gapDelta} to first obtain $\lambda_{\textrm{SO}}$ from the value of the \textit{ab-initio} topological gap at null electric field. Then, we apply a finite external electric field to the system with the methodology explained in the previous paragraph, and use \ref{eq:gapDelta} to deduce the total effective field $E^{\textrm{tot}}_z$ associated to the \textit{ab-initio} energy gap that arises at such given external electric field. 

The electron-phonon coupling is computed in absence of electric-field, with the same grids used for the Born Effective charges calculation. In detail, we compute $\langle g^2_{\pmb{\Gamma}} \rangle=\sum_{i,j=c,v}\Big{|}\langle \psi_{i\mathbf{K}} |\frac{\partial H}{\partial \mathbf{u}_{\Gamma, \nu}} |\psi_{j\mathbf{K}} \rangle \cdot \mathbf{e}^{\nu}/\sqrt{2M\omega_{\Gamma, \nu}/\hbar}\Big{|}^2/4$, where $\psi_{i/j,\mathbf{K}}$ are the valence/ conduction (v/c) Bloch functions at the K point, and we then mediate over the degenerate modes $\nu=$TO/LO as well. The parameter used in the tight-binding method are found by $\xi=\frac{1}{b_0t_1}\sqrt{\frac{4 \langle D^2_{\pmb{\Gamma}} \rangle}{9}}$, where $\langle D^2_{\pmb{\Gamma}} \rangle=\langle g^2_{\pmb{\Gamma}} \rangle (2M\omega_{\Gamma}/\hbar)$ and $t_1$ is fitted from the first principles bands and $b_0$ is the nearest neighbour distance, as detailed in Refs.\cite{PhysRevB.84.035433, Bistoni2019}.  For germanene, we obtain $\xi=0.394 \frac{1}{\text{\AA } ^2}$ .

\textit{Symmetries of Jacutingaite atoms---} Hg atoms occupy the 2d Wyckoff Position (WP), like the Ge atoms in Germanene, so the symmetries constrain the corresponding Born effective charge tensors to be diagonal with two independent components that are equal on equivalent Hg atoms in the $p\bar{3}m1$ phase, being related by inversion symmetry.
Pt atoms occupy two different WPs, $1a$ and $3e$. The $1a$ platinum sitting at the center of each honeycomb, henceforth labelled as Pt$_1$, displays the full crystallographic point-group symmetry $D_{3d}$ ($\bar{3}m$) and a diagonal charge tensor with equal in-plane components. The $3e$ Pt atoms, which contribute to the MLWFs realizing the Kane-Mele effective model and henceforth labeled as Pt$_2$, display a $2/m$ site-symmetry, being related by the threefold rotation about the out-of-plane direction. Finally, Se atoms occupy the $6i$ WP, with $m$ site symmetry and being related by inversion and threefold rotation symmetries. The Born effective charge tensor has 5 independent components for both Pt$_2$ and Se atoms.  Considering the Pt$_2$ and Se atoms for which the twofold rotation axis is parallel to $x$, site symmetries impose $Z^*_{\textrm{Pt}_{2}/\textrm{Se},xy}=Z^*_{\textrm{Pt}_{2}/\textrm{Se},xz}=Z^*_{\textrm{Pt}_{2}/\textrm{Se},zx}=Z^*_{\textrm{Pt}_{2}/\textrm{Se},yx}=0$.
\clearpage
\nocite{PhysRevLett.77.3865,VANSETTEN201839,PhysRevB.88.085117, PhysRevB.109.075420, PhysRevB.13.5188,PhysRevB.96.075448, Giannozzi2017, PhysRevLett.77.3865,PhysRevLett.93.185503,PhysRevB.84.035433, Pizzi2020}
\printbibliography
\end{document}